\documentclass[pra,showkeys,showpacs,groupedaddress,twocolumn]{revtex4}
\usepackage{amssymb}
\usepackage{graphicx}
\usepackage{dcolumn}
\usepackage{bm}
\usepackage{amsmath}
\usepackage{subfigure}
\usepackage{float}
\usepackage{color}

\begin{document}
\title{Transition from electromagnetically induced transparency to Autler-Townes splitting in cold cesium atoms}

\author{Liping Hao$^{1}$ }
\author{Yuechun Jiao$^{1,2}$}
\author{Yongmei Xue$^{1}$}
\author{Xiaoxuan Han$^{1}$}
\author{Suying Bai$^{1}$}
\author{Jianming Zhao$^{1,2}$}
\thanks{Corresponding author: zhaojm@sxu.edu.cn}
\author{Georg Raithel$^{1,3}$}
\affiliation{$^{1}$State Key Laboratory of Quantum Optics and Quantum Optics Devices, Laser spectroscopy Laboratory, Shanxi University, Taiyuan 030006, P. R. China}
\affiliation{$^{2}$Collaborative Innovation Center of Extreme Optics, Shanxi University, Taiyuan
030006, China}
\affiliation{$^{3}$ Department of Physics, University of Michigan, Ann Arbor, Michigan 48109-1120, USA}

\date{\today}

\begin{abstract}
Electromagnetically induced transparency (EIT) and Aulter-Townes splitting (ATS) are two similar yet distinct phenomena that modify the transmission of a weak probe field through an absorption medium in the presence of a coupling field, featured in a variety of three-level atomic systems. In many applications it is important to distinguish EIT from ATS splitting. We present EIT and ATS spectra in a cold-atom three-level cascade system, involving the 35$S_{1/2}$ Rydberg state of cesium.
The EIT linewidth, $\gamma_{EIT}$, defined as the full width at half maximum (FWHM), and the ATS splitting, $\gamma_{ATS}$, defined as the peak-to-peak distance between AT peak pairs, are used to delineate the EIT and ATS regimes and to characterize the transition between the regimes. In the cold-atom medium, in the weak-coupler (EIT) regime $\gamma_{EIT}$ $\approx$ A + B($\Omega_{c}^2$ + $\Omega_{p}^2)/\Gamma_{eg}$, where $\Omega_{c}$ and $\Omega_{p}$ are the coupler and probe Rabi frequencies, $\Gamma_{eg}$ is the spontaneous decay rate of the intermediate 6P$_{3/2}$ level, and parameters $A$ and $B$ that depend on the laser linewidth. We explore the transition into the strong-coupler (ATS) regime, which is characterized by the linear relation $\gamma_{ATS}$ $\approx$ $\Omega_{c}$. The experiments are in agreement with numerical solutions of the Master equation.
\end{abstract}
\maketitle

\section{Introduction}
Electromagnetically induced transparency (EIT)~\cite{Boller} is a quantum interference effect in which the absorption of a weak probe laser, interacting resonantly with an atomic transition, is reduced in the presence of a coupling laser.
EIT is, for instance, crucial in optically controlled slowing of light~\cite{Hau} and optical storage \cite{Phillips}.
Aulter-Townes splitting (ATS)~\cite{Autler}, a linear (resonant) AC Stark effect proposed by Aulter and Townes, was observed originally in the microwave and later the light domain.
EIT and ATS have been extensively investigated experimentally and theoretically in $\Lambda$-, $V-$ and cascade-type three-level atoms~\cite{Boller,Xiao,Scully,Fleischhauer,Anisimov,Abi-Salloum}. Holloway $et$ $al.$~\cite{holloway2017} have investigated the relationship between the Rabi frequency of resonant RF transitions between Rydberg states and the resultant ATS splitting in Rydberg-EIT spectra measured in room-temperature atomic vapor.
While EIT and ATS may phenomenologically look similar, they are different in nature, leading to an interest in the establishment of criteria to discern them. Anisimov $et$ $al.$~\cite{Anisimov} propose an objective method based on Akaike's information criterion, to discern ATS from EIT in experimental data of a $\Lambda$-type three-level atom, also applicable to an upper-level-driven $\Xi$-type case.
In~\cite{Abi-Salloum} a threshold between EIT and ATS, $\Omega_{t}$, is introduced that is defined via polarization decay rates. According to this threshold, ATS (two resonances with a gap in between) is observed in four different three-level systems in a strong-coupling-field regime ($\Omega_{c}$ / $\Omega_{t}$ $>$ 1). EIT (transparency due to destructive interference) is observed only in $\Lambda$- and cascade-EIT configurations in a weak-coupling-field regime ($\Omega_{c}$ / $\Omega_{t}$ $<$ 1). Both Refs.~\cite{Anisimov, Abi-Salloum} are for systems that are free of Doppler effects.

In the present experimental and theoretical study, we focus on cold-atom cascade EIT and ATS in a magneto-optical trap (MOT), with a Rydberg upper-level state. Cold-atom clouds exhibit a cleaner transition between EIT and ATS than room-temperature atomic vapor because cold atoms are often free of significant Doppler mismatch between coupler and probe fields. In contrast, the Doppler mismatch strongly affects linewidths in vapor-cell EIT and ATS spectra~\cite{holloway2017}. Our analysis, which accounts for any residual Doppler mismatch, leads to readily accessible criteria to distinguish cold-atom EIT from ATS. In our work we use a variation of cascade systems, Rydberg-EIT, that
was first observed in a vapor cell~\cite{Mohapatra} and later in a rubidium MOT~\cite{Pritchard}.
Rydberg-EIT has been used to realize a single-photon transistor~\cite{Gorniaczyk} and a single-photon source~\cite{Dudin,Viscor} by employing the blockade effect~\cite{Singer,Tong,Vogt}, which results from strong
interactions between Rydberg atoms. ATS involving Rydberg atoms has also been investigated in rubidium~\cite{Teo,DeSalvo} and cesium~\cite{Zhang,Zhang2014}. Here we study the dependence of the Rydberg-EIT linewidth and the ATS in a cesium MOT on the Rabi frequencies of the probe and coupling transitions.

\section{Theoretical model}

We consider the cesium  cascade three-level system shown in Fig.~1(a). The coupling laser drives the upper transition, $|6P_{3/2}, $F'$ = 5 \rangle$ ($|2\rangle$) $\rightarrow$ $|35S_{1/2}\rangle$ ($|3\rangle$). The weak probe laser couples the lower transition, $|6S_{1/2}, $F$ = 4\rangle$ ($|1\rangle$) $\rightarrow$ $|6P_{3/2}, $F'$ = 5 \rangle$ ($|2\rangle$).
The respective wavelengths and Rabi frequencies are $\lambda_c$ and $\Omega_c$, and $\lambda_p$ and $\Omega_p$. In the rotating-wave approximation and the field picture, the Hamiltonian of the three-level atom represented in the space $\{ \vert 1 \rangle \, , \vert 2 \rangle \, , \vert 3 \rangle \}$ is

\begin{figure}[htbp]
\vspace{1ex}
\centering
\includegraphics[width=0.5\textwidth]{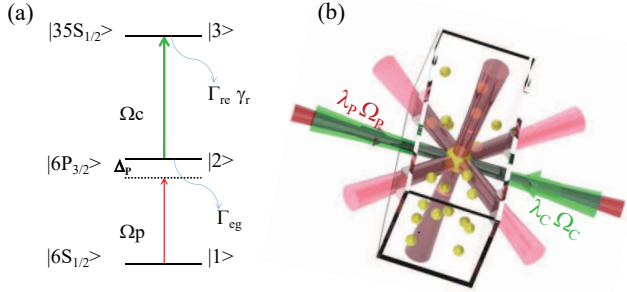}
\vspace{-3ex}
\caption{(color online) (a) Scheme of the cascade three-level atom. The coupling laser is resonant with the Rydberg transition, $|6P_{3/2}, $F'$ = 5 \rangle$ ($|2\rangle$) $\to$  $|35S_{1/2}\rangle$ ($|3\rangle$) (wavelength 510~nm, Rabi frequency $\Omega_{c}$). The weak probe beam (wavelength 852~nm, Rabi frequency $\Omega_{p}$) is referenced to the transition $|6S_{1/2}, F = 4\rangle$ ($|1\rangle$) $\to$ $|6P_{3/2}, $F'$ = 5\rangle$ ($|2\rangle$) using a Doppler-free polarization-spectroscopy setup, and is scanned over the resonance. (b) Sketch of the experimental setup. The laser beams are separated via a dichroic mirror (not shown), and the probe light is detected using a single-photon counter module (SPCM).}
\end{figure}

\begin{equation}
H=\frac{\hbar}{2}\left(\begin{array}{ccc}
0&\Omega_{p}&0\\
\Omega_{p}&-2\Delta_{p}&\Omega_{c}\\
0&\Omega_{c}&-2(\Delta_{p}+\Delta_{c})
\end{array}\right), \label{Hmatrix}
\end{equation}
where $\Delta_{c}$ and $\Delta_{p}$ are the detunings of the coupling and probe beams, respectively.
To account for decay and dephasing, the system is described using the Lindblad equation for the density matrix $\rho$,
\begin{equation}
\dot{\rho} = - \frac{i}{\hbar}{[H,\rho]+\mathfrak{L}}
\end{equation}
where $\mathfrak{L}$ is the Lindblad operator that accounts for
the decay processes in the atom. In the space $\{ \vert 1 \rangle \, , \vert 2 \rangle \, , \vert 3 \rangle \}$, $\mathfrak{L}$  becomes~\cite{Raitzsch}

\begin{equation}
\mathfrak{L}=\left(\begin{array}{ccc}
\Gamma_{eg}\rho_{22}&-\frac{1}{2}\gamma_{2}\rho_{12}&-\frac{1}{2}\gamma_{3}\rho_{13}\\
-\frac{1}{2}\gamma_{2}\rho_{21}&-\Gamma_{eg}\rho_{22}+\Gamma_{re}\rho_{33}&-\frac{1}{2}(\gamma_{2}+
\gamma_{3})\rho_{23}\\
-\frac{1}{2}\gamma_{3}\rho_{31}&-\frac{1}{2}(\gamma_{2}+\gamma_{3})\rho_{32}&-\Gamma_{re}\rho_{33}
\end{array}\right), \label{gamma}
\end{equation}

where $\gamma_{2}$ and $\gamma_{3}$ are the dephasing rates of the intermediate and Rydberg states, respectively. It is $\gamma_{2}$ = $\gamma_{e}$ + $\Gamma_{eg}$, where $\gamma_{e}$ is a collision-induced dephasing rate of level $|2\rangle$, and $\gamma_{e} \ll \Gamma_{eg} = 2\pi \times$~5.2~MHz. Further,
$\gamma_{3}$ = $\gamma_{r}$ + $\Gamma_{re}$. For the Rydberg level the population decay rate $\Gamma_{re}$ is, typically, smaller than the dephasing $\gamma_{r}$, because Rydberg-atom lifetimes are long (lifetimes are $\sim n^3$ and on the order of 100~$\mu$s) and interactions between Rydberg atoms in cold-atom clouds are often strong (van der Waals interactions scale as $n^{11}$).

The spectrum is given by the probe-power transmission, $P = P_0 \exp(-\alpha L)$, with the probe-laser absorption coefficient, $\alpha$ = 2$\pi$Im($\chi$)$/$$\lambda_p$, the MOT size, $L$, and the susceptibility of the medium seen by the probe laser, $\chi$. The susceptibility, $\chi$, is

\begin{equation}
\chi = \frac{2N\mu_{12}}{E_p\epsilon_0} \rho_{12},
\end{equation}
where $N$ is the average atomic density, $\mu_{12}$ is the dipole moment of transition $|1\rangle$ $\to$ $|2\rangle$, $E_p$ is the amplitude of the probe, $\epsilon_0$ is the vacuum permittivity,
and $\rho_{12}$ is the density matrix element between $|1\rangle$ and $|2\rangle$.

We numerically solve the Eqs.~(1)-(3) to obtain the steady-state absorption coefficient $\alpha$ for a range of values of $\Omega_{c}$ and $\Omega_{p}$. The result is averaged over the thermal velocity distribution in the gas~\cite{holloway2017}; under our conditions ($T= 100$ to $200~\mu$K) the thermal motion is not very important. In the calculation we also assume $\gamma_r=0$, which is admissible due to our very low experimental probe intensities and the low principal quantum number of the utilized Rydberg state.
In Fig.~2(a) and (b) we show $\alpha$  as a function of $\Omega_{c}$ and $\Delta_{p}$, for fixed $\Omega_{p} = 2 \pi \times$~1.05~MHz and $\Delta_c = 0$. The left spectrum in Fig.~2(a) is in the EIT regime, where $\Omega_p \lesssim \Omega_{c} = 2 \pi~\times~2.03$~MHz $< \Gamma_{eg}$. In the EIT case,  the width of the EIT window is most appropriately described by the full width at half maximum (FWHM) of the dip in absorption, $\gamma_{EIT}$, which fills a small fraction of the natural linewidth of the probe transition. The right curve in Fig.~2(a) is for $\Omega_{c} = 2 \pi~\times$~13.01~MHz, which is  $> \Gamma_{eg}$ and therefore in the Autler-Townes regime. In that regime, the behavior is more appropriately described by the peak-to-peak spacing of the line pair, $\gamma_{ATS}$. As seen in Figs.~2(a) and (b), in the weak-probe limit and for $\Delta_c = 0$, in the AT regime the peaks tend to have a width of $\Gamma/2 = 2 \pi \times 2.6$~MHz, as the Rydberg states are very long-lived ($\Gamma_{re}$ $\ll$ $\Gamma_{eg}$) and do not contribute to the width of the AT peaks, and the spacing $\gamma_{ATS} \approx \Omega_c$.

In Fig.~2(c) we show the calculated $\gamma_{EIT}$ as a function of $\Omega_{c}$ and $\Omega_{p}$. In the EIT domain, outlined by the dashed quarter-circle in Fig.~2(c), $\gamma_{EIT} \approx (\Omega_{c}^2 + \Omega_{p}^2)/\Gamma$. It is noted that in hot gases $\gamma_{EIT}$ tends to be much larger, in particular in weak coupler and probe fields, and follows a different scaling. This is due to the difference between coupler and probe wavelengths and the generally large size of the Doppler shifts (see~\cite{holloway2017} and references therein). Figure~2(c) also shows that in the AT regime, characterized by $\Omega_{c} \gtrsim \Gamma_{eg}$, it is $\gamma_{ATS} \approx \Omega_{c}$, largely independent of $\Omega_p$.
The transition between the EIT (weak coupler) and ATS (strong coupler) regimes occurs at $\Omega_c \sim \Gamma_{ge}$. Similar calculations can be performed at any atom temperature~\cite{holloway2017}, and for both cascade and $\Lambda$-type three-level atoms. In the hot-atom cases, the difference between coupler and probe wavelengths typically determines the visibilities and detailed shapes of EIT and ATS spectra.

\begin{figure}[htbp]
\vspace{1ex}
\centering
\includegraphics[width=0.35\textwidth]{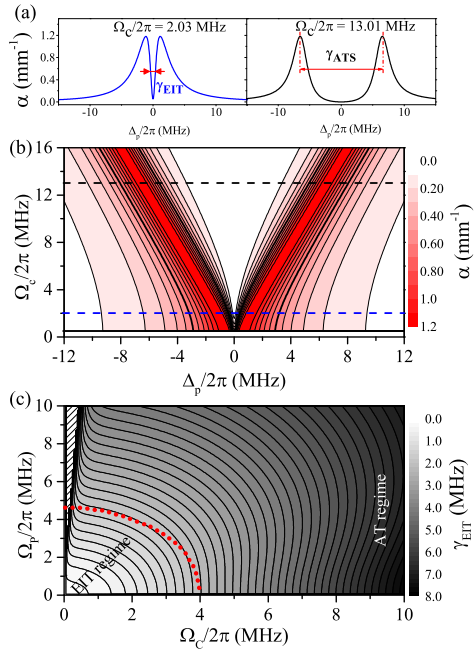}
\vspace{-3ex}
\caption{(color online) (a) Calculations of the probe absorption coefficient of cold cesium atoms in a MOT, $\alpha$, for $\Omega$$_{p}$ = 2$\pi$~$\times$~1.05~MHz and $\Omega_{c}$ = 2$\pi$~$\times$~2.03~MHz (left) and
2$\pi$~$\times$~13.01~MHz (right), atom density $10^{10}$cm$^{-3}$ and vanishing laser line width.
The FHWM of the EIT window, $\gamma_{EIT}$, and the ATS, $\gamma_{ATS}$, are defined as shown. (b) Same calculation as in (a) for
$\Omega_{c}$ ranging from $2 \pi \times$0.5~MHz to $2 \pi \times$16~MHz. The cuts along the horizontal
dashed lines correspond to the spectra shown in (a).
(c) Calculated $\gamma_{EIT}$ as a function of $\Omega_{p}$ and $\Omega_{c}$. Within the displayed parameter space, the EIT regime is found in the lower-left corner. In that regime, $\gamma_{EIT} \approx (\Omega_{c}^2 +
\Omega_{p}^2)/\Gamma_{eg} \lesssim \Gamma_{eg}$, {\sl i.e.} $\gamma_{EIT}$ is approximately proportional to the sum of the squares of the Rabi frequencies. In the strong-coupler limit, $\Omega_{c} > \Gamma_{eg}$, both $\gamma_{EIT}$  and $\gamma_{ATS}$ follow a linear dependence on $\Omega_{c}$ and $\gamma_{ATS} \approx \Omega_{c}$. This is the Autler-Townes regime.}
\end{figure}

\section{Experimental measurement}

The EIT and ATS experiments are performed in a magneto-optical trap (MOT) with temperature $\sim$ 100~$\mu$K and atomic density $\sim$ 10$^{10}$~cm$^{-3}$. The coupling and probe lasers have linear and parallel polarizations and counter-propagate through the cold-atom cloud, as seen in Fig.~1(b). The details of the experiment have been described previously~\cite{Jiao2016}.
The probe beam is derived from a diode laser (DLpro, Toptica) that is locked to the ground-state transition, $|1\rangle$ $\to$ $|2\rangle$, using polarization spectroscopy~\cite{Pearman}; the beam has a Gaussian waist $\omega_{p0}$ = 10~$\mu$m at the MOT center.
The strong coupling laser (Toptica TA-SHG110) has a Gaussian waist $\omega_{c0}$ = 30~$\mu$m and drives the Rydberg transition $|2\rangle$ $\to$ $|3\rangle$. The frequency of the coupling laser is stabilized to the Rydberg transition using a Rydberg-EIT signal obtained from a cesium room-temperature vapor cell~\cite{Jiao}. In each experimental cycle, after turning off the trap beams, we switch on the coupling and probe lasers for 25~$\mu$s. During the probe pulse the probe-laser frequency is swept across the $|6S_{1/2},F=4\rangle$ $\to$ $|6P_{3/2},F=5\rangle$ transition using a double-pass acousto-optic modulator (AOM) over a range of $\pm$10~MHz relative to the transition center.

In order to avoid Rydberg excitation blockade and interaction effects, and to be able to reach high coupling Rabi frequencies, $\Omega_c$, we have chosen a low principal quantum number $n$ (35$S_{1/2}$).
Further, to ensure that radiation pressure~\cite{Jiao2016,radiation} has a negligible effect on the EIT and ATS spectral profiles, we use a very low probe power,  $P_{852}$ = 200~pW, and a single-photon counter module (SPCM) for probe-light detection. Under these conditions, the radiation-pressure-induced velocity change during the probe pulse has an upper limit of $\approx 7~$cm/s, corresponding to Doppler shifts $<100~$kHz (which is negligible in the present work). The EIT and ATS spectra are recorded using a data acquisition card (NI-PCI-6542) and processed with a Labview program.

\begin{figure}[htbp]
\vspace{1ex}
\centering
\includegraphics[width=0.35\textwidth]{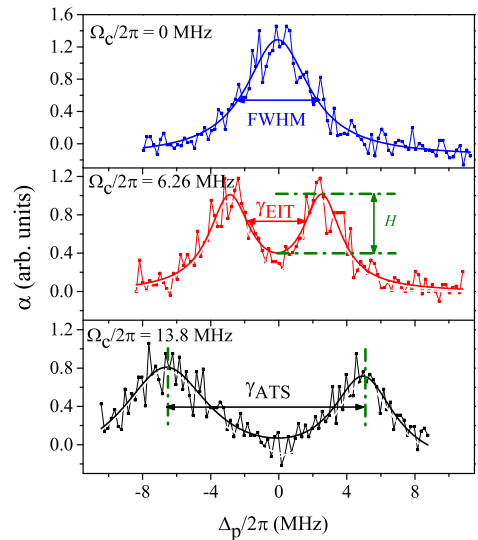}
\vspace{-3ex}
\caption{(color online) Measurements of EIT and ATS absorption spectra for $\Omega$$_{p}$ = 2$\pi$~$\times$~1.05~MHz and the indicated Rabi frequencies of the coupling laser, $\Omega_{c}$ = 2$\pi$~$\times$~0~MHz (top), 6.26~MHz (middle) and 13.8~MHz (bottom). The solid lines show the results of Lorentzian multi-peak fits. The coupler-free linewith (top curve) is 2$\pi \times$ (4.53 $\pm$ 0.29)~MHz, which is close to the expected value of 2$\pi\times$~$5.2$~MHz.
The FWHM EIT linewidth, $\gamma_{EIT}$, and the ATS, $\gamma_{ATS}$, are obtained from the fit functions, as indicated. The depth of the EIT dip (or, in the ATS regime, the depth of the valley between the AT peaks), $H$, is defined as the difference between the peak absorption, averaged over the two peaks, and the absorption minimum between the peaks.}
\end{figure}

\begin{figure}[htbp]
\vspace{1ex}
\centering
\includegraphics[width=0.45\textwidth]{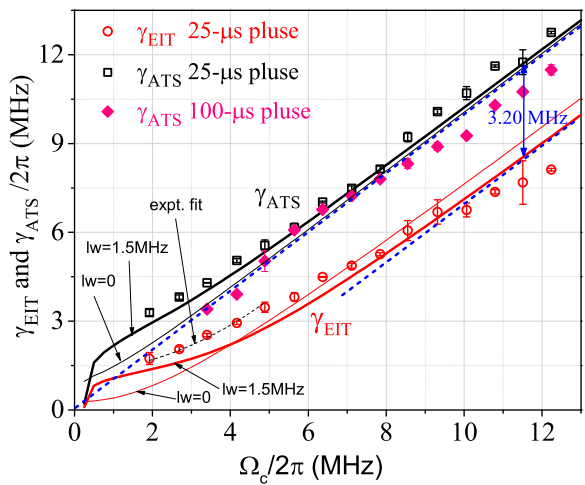}
\vspace{-3ex}
\caption{Measurements (symbols) and calculations (solid lines) of the $\gamma$$_{ATS}$ and $\gamma$$_{EIT}$ as a function of $\Omega$$_{c}$ for $\Omega$$_{p}$ = 2$\pi$~$\times$~1.05~MHz and probe/coupling duration 25~$\mu$s (hollow) and 100~$\mu$s (filled). In the calculation, the laser linewidth is 2$\pi$~$\times$ 1.5~MHz (bold solid lines) and 0 (thin solid lines).
The black thin dashed line shows a fit to the experimental data; the fit function is $\gamma_{EIT} \approx A + B (\Omega_{c}^2 +\Omega_{p}^2)/\Gamma_{eg}$ with $A$ = 2$\pi \times$ (1.42 $\pm$ 0.10)~MHz and $B$ = 0.44 $\pm$ 0.03. }
\end{figure}

The absorption coefficient, $\alpha$, is obtained from the measured SPCM count number, $P(\Delta_p)$, and the off-resonant (absorption-free) count number, $P_0$, using the relation $\alpha L = -\ln (P(\Delta_p)/P_0)$, where $L$ is the effective MOT diameter along the probe beam path. In Fig.~3, we present EIT and ATS sample spectra that explain our linewidth parameters. The EIT linewidth, $\gamma_{EIT}$, is defined as the full width at half maximum (FWHM) of the absorption minimum measured as a function of probe-laser frequency, while the ATS, $\gamma_{ATS}$, is defined as the distance between the centers of the two
AT absorption peaks. Using these definitions, we have determined $\gamma_{EIT}$ and $\gamma_{ATS}$ over a range of the coupling Rabi frequency $\Omega_{c}$, for fixed $\Omega_p$. In Fig.~4 we show the measured (hollow symbols) and calculated (lines) results for $\gamma_{EIT}$ and $\gamma_{ATS}$ versus $\Omega_{c}$. The calculated values are obtained in a way analogous
to Fig.~3. To account for laser frequency jitter, the calculated spectra are convoluted with a Gaussian of 1.5~MHz FWHM before determination of $\gamma_{EIT}$ and $\gamma_{ATS}$ (also see discussion of Fig.~5 below).

We first discuss $\gamma_{ATS}$. In the AT regime, $\Omega_c \gtrsim \Gamma_{eg}$, it is $\gamma_{ATS} \approx \Omega_c$ (see dashed line in Fig.~4). In the EIT regime, $\Omega_c \lesssim \Gamma_{eg}$, the peak-to-peak splitting $\gamma_{ATS} > \Omega_c$. The spectral broadening due to the laser linewidth generally enhances this trend, because the outer wings of the split lines are wider than the inside wings, hence the spectral averaging pulls the line centers outward (see Fig.~2). Due to the laser linewidth and other broadening mechanisms, the absorption minimum at $\Delta_p = 0$ disappears entirely when $\Omega_c$ drops below a critical value, which is $\approx 2 \pi \times 2$~MHz in our case. A similar nonlinear behavior of AT splitting is observed in Ref.~\cite{holloway2017}, where an RF-induced ATS is used to measure a microwave electric field. Here, the calculations for 1.5~MHz FWHM laser linewidth (bold solid lines in Fig.~4) reproduce the measurements best.

We also show a similar measurement with 100-$\mu$s probe- and coupler-pulse sweeps (filled squares in Fig.~4). It is seen that $\gamma_{ATS}$ for the longer pulses is less than in the 25-$\mu$s case. We attribute the reduction in $\gamma_{ATS}$ in part to
dephasing caused by Rydberg-atom interactions~\cite{Zhang2014}. Longer pulses will generally lead to a higher Rydberg-atom number in the atom-field interaction volume, causing dephasing by Rydberg-Rydberg interactions. Longer times will also increase the likelihood of Penning and thermal ionization. Any ions in the sample would contribute to Rydberg-level dephasing via the ion electric fields.  Also, for the long pulses radiation pressure will have an enhanced broadening effect that tends to reduce $\gamma_{ATS}$ (up to several $100$~kHz).

In the EIT domain, $\Omega_{c} \lesssim \Gamma_{eg}$, the EIT linewidth $\gamma_{EIT}$ in Fig.~4 exhibits a quadratic behavior, as expected from Fig.~2(c). Here, we find that in the EIT domain the experimental data are fit quite well by an equation $\gamma_{EIT} \approx A + B (\Omega_{c}^2 +\Omega_{p}^2)/\Gamma_{eg}$, as shown by the black dashed line. The fitting parameters are $A = 2\pi \times (1.42 \pm 0.10)$~MHz and $B = 0.44 \pm 0.03$. The laser-linewidth-induced line broadening of the EIT spectra reduces the pre-factor $B$ to a value significantly below its ideal value of $\approx 1$, and it adds an additive constant on the order of the laser linewidth. In the AT regime, $\Omega_{c} \gtrsim \Gamma_{eg}$, the measured FWHM width of the spectral window of reduced absorption follows a trend $\gamma_{EIT} = \Omega_c - 2 \pi \times 3.2$~MHz, which agrees well with the calculation for 1.5~MHz FWHM laser linewidth. In the absence of laser-line broadening it would be $\gamma_{EIT} = \Omega_c - \Gamma_{eg}/2$  $= \Omega_c - 2 \pi \times 2.6$~MHz.

Figures 2 and 4 clearly demonstrate the difference between EIT and ATS. In the EIT regime the value of $\gamma_{EIT}$ has more physical meaning, whereas in the AT regime the value of $\gamma_{ATS}$ has more physical meaning.
In the EIT regime, $\Omega_{c} \lesssim \Gamma_{eg}$, it is $\gamma_{EIT}$ $\simeq$ A + B($\Omega_{c}^2 + \Omega_{p}^2$)/$\Gamma_{eg}$, where for vanishing laser linewidth $A \rightarrow 0$ and $B \rightarrow 1$. In the AT regime, $\Omega_{c} \gtrsim \Gamma_{eg}$, it is $\gamma_{ATS}$ $\simeq$ $\Omega$$_{c}$. The individual AT-lines have a minimal width of $\Gamma_{eg}/2$. Convolution with the laser line profile leads to some additional broadening. The MOT magnetic field, which is always left on here, may also contribute to the line broadening. In the case of $\Omega_p$ approaching and exceeding $\Gamma_{eg}$, the AT lines also become saturation-broadened (this effect is negligible in our case because $\Omega_p$ is only $2 \pi \times 1.05$~MHz). The more the AT lines are broadened by these effects, the more the difference $\gamma_{ATS} - \gamma_{EIT}$ exceeds its lower limit of $\Gamma_{eg}/2$.

We also note that $\Omega_c$ has a different significance in the two domains. In the ATS regime $\Omega_c$ splits the lines but does not broaden them, whereas in the EIT regime $\Omega_c$ broadens the EIT linewidth ($\gamma_{EIT}$ increases with $(\Omega_{c}^2 + \Omega_{p}^2)$). Conversely, laser linewidth and other broadening mechanisms also have a different significance in the two domains. In the EIT domain the broadening affects the parameters in the fit function $\gamma_{EIT}$ $\simeq$ A + B($\Omega_{c}^2 + \Omega_{p}^2$)/$\Gamma_{eg}$ ($A=0$ and $B=1$ in the absence of broadening). In the ATS domain the broadening has comparatively little effect on $\gamma_{ATS}$.

\begin{figure}[htbp]
\vspace{1ex}
\centering
\includegraphics[width=0.4\textwidth]{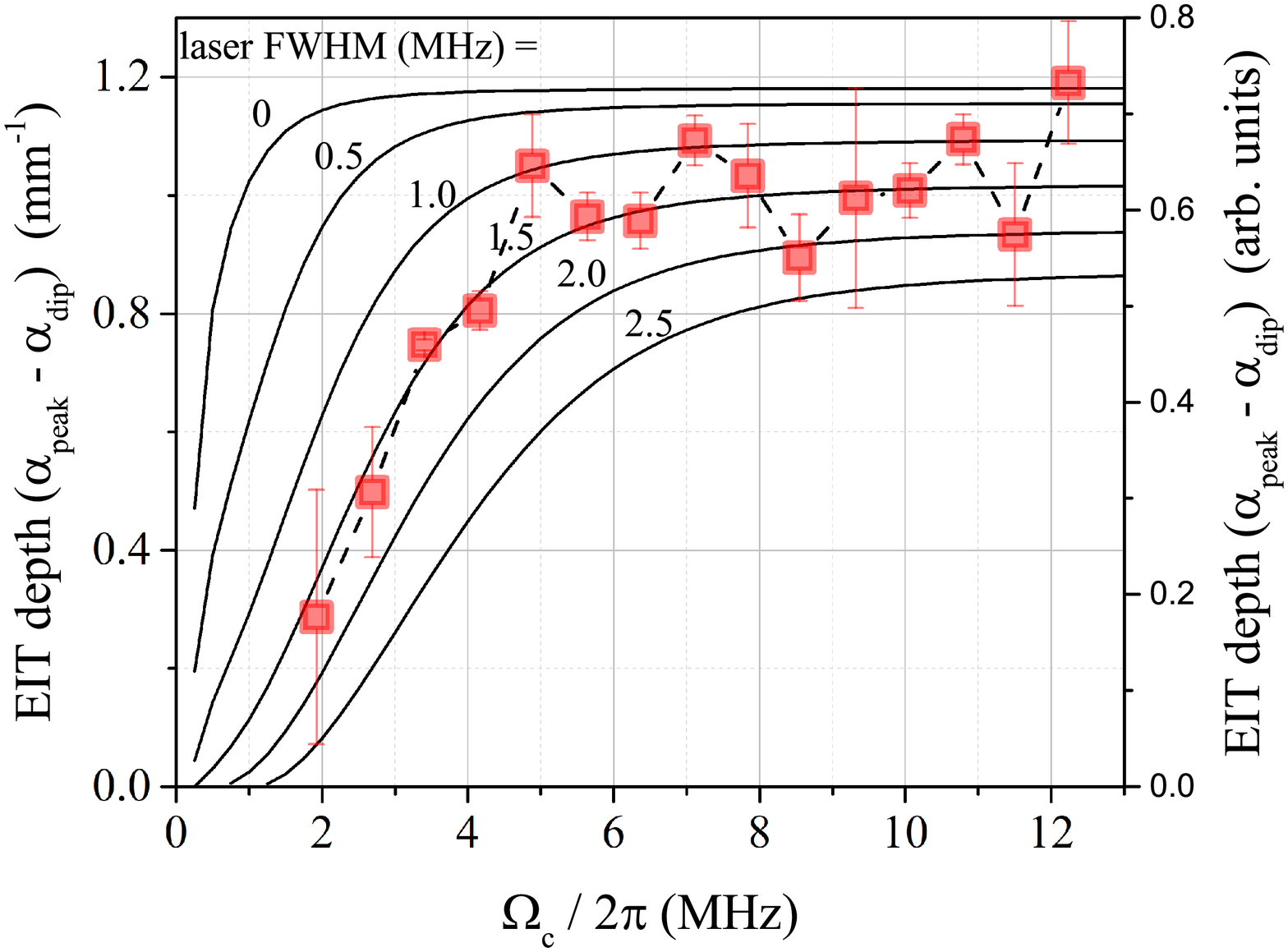}
\vspace{-3ex}
\caption{Measurements (symbols) and calculations (black solid lines) of the EIT depth, $H$, defined in Fig.~3, vs $\Omega_{c}$ for $\Omega_{p} = 2 \pi~\times$~1.05~MHz. The upper atomic level is the 35$S_{1/2}$ state. The FWHM laser linewidths in the calculations range from 0 to 2.5~MHz. The line for 1.5~MHz laser linewidth reproduces the experimental data best.}
\end{figure}

An important measure for any practical application of EIT is the depth of the EIT line, $H$, defined as the difference between the absorption coefficients on the peaks and at the center of the EIT dip (see middle curve in Fig.~3). The depth $H$ increases with $\Omega_{c}$ in the EIT regime and plateaus at a laser-linewidth-dependent maximal value in the ATS regime, as shown by the calculated curves in Fig.~5. Assuming a FWHM laser linewidth of 1.5~MHz, we achieve a good agreement between experiment and calculation. This best-fit laser linewidth agrees well with manufacturer estimates for the utilized laser systems.

\section{Conclusion}

We have presented measured and calculated cold-atom EIT and AT absorption spectra of a cascade three-level atom involving the 35$S_{1/2}$ Rydberg state. The measurements show good agreement with calculations. The spectra exhibit two regimes, an EIT regime for weak $\Omega_{c}$ and an ATS regime for large $\Omega_{c}$.
While similar in appearance, the EIT and AT spectra are different in physical interpretation~\cite{Anisimov,Abi-Salloum}. We have defined widths $\gamma_{EIT}$ and $\gamma_{ATS}$ that help making the distinction between EIT and ATS. We find $\gamma_{EIT} = A+B((\Omega_{c}^2 + \Omega_{p}^2)/\Gamma_{eg})$ when $\Omega_{c} \lesssim \Gamma_{eg}$ (the cold-atom EIT regime); the parameters $A$ and $B$ depend on broadening  due to laser linewidth etc.
In the AT regime, $\Omega_{c} \gtrsim \Gamma_{eg}$, it is $\gamma_{ATS} = \Omega_{c}$ and $\gamma_{EIT} = \gamma_{ATS} - C$, with a constant $C \gtrsim \Gamma_{eg}/2$; the lower limit of $C$ is realized in the absence of broadening due to laser linewidth etc.

Clear criteria to distinguish between EIT and ATS are valuable in a wide variety of atomic-physics, quantum-optics and quantum information applications of these schemes. The distinction of EIT from ATS is generally important in understanding the quantum physics of atom-light interaction. The details of ATS splitting and line broadening effects are
especially important in applications that deal with quantitative, atom-based microwave field measurements using
Rydberg-EIT and microwave-coupled ATS~\cite{Sedlacek2012,holloway2017}. Cold-atom, narrow-linewidth EIT and ATS are helpful in improving the accuracy and resolution of the atom-based field measurements.

The work was supported by the National Key R\&D Program of China (Grant No. 2017YFA0304203), the National Natural Science Foundation of China (Grants No. 61475090, and No. 61675123, and No. 61775124), Changjiang Scholars and Innovative Research Team in University of Ministry of Education of China (Grant No. IRT13076), and the State Key Program of National Natural Science of China (Grant No. 11434007). GR acknowledges support by the NSF (PHY-1506093) and BAIREN plan of Shanxi province.

%\section{References}

\end{document}